# On some extreme cohesive properties of alpha-Plutonium: temperature dependence of elastic moduli, large pressure derivative of bulk modulus and Grūneisen parameter

S.K. Sikka

Scientific Consultant to the office of the Principal Scientific Adviser to Government of India, Vigyan Bhawan Annexe, Maulana Azad Road, NewDelhi-110011

## Abstract

The basic observed cohesive properties of $\alpha$-Pu like temperature variation of elastic moduli, large value of the pressure derivative of the bulk modulus and anomalous values of the Gruneisen parameter in literature are not yet completely understood. In this communication, we provide new insights into these.

Plutonium (Z=94) is a man-made element. Many of its isotopes are fissionable. Therefore, it was first used in implosion atomic bombs and then in nuclear reactors and other devices for power generation. Pu belongs to the actinide series, which contain 5f electrons. In the earlier part of the series, the 5f electrons are delocalized and behave like the 5d electrons of the transition metals. The actinides from Am onwards have localised 5f electrons. Pu lies at boundary of the delocalized and localized 5f electrons. Right at Pu, between the room temperature monoclinic α phase and the high temperature face δ-phase, there appears to be a transition in the 5f electronic nature from bonding (partial delocalisation) to partial localisation. This is responsible for its extreme physical, chemical and mechanical properties. These are now well documented [1]. Some of these of relevance to some cohesive properties are not yet completely understood. These are as follows :

The low temperature α–Pu undergoes five structural transitions ,β at 398 K, γ at 488 K, δ at 593 K, δ' at 736 K and ε at 756 K up to the melting point of 913 K with significant volume expansions. The volume expands by about 25% between the α and the δ phases. The volume contracts for the δ and δ' phases [2]. A seventh phase occurs at a pressure of 37 GPa[3],where the α–Pu, a 16-atom /unit cell monoclinic structure undergoes transformation to a 4 -atom/unit cell distorted α–u structure in Pnma space group[4]. Theoretically, it has been shown that these transformations are manifestations of varying degrees of 5f correlations. For example, using GW approach (G=solid Green function, W=dynamically screened electron - electron interaction), Savane et al [5] have defined a correlation strength (C= 1- $w_{rel}$) where $w_{rel}$ is the relative f- band width reduction in GW compared to local density approximation (LDA) of the density functional theory and successfully explained physical properties like atomic volumes (essentially thermal expansion), sound speeds and resistivity of these Pu phases.

The temperature dependence of the elastic moduli of different phases have recently been remeasured employing the resonant-ultrasound spectroscopy technique [6-9] .In particular, Fig.1 shows the variation of bulk moduli (B) and shear moduli (G) for the α-phase. The bulk modulus ratio between 0 and 300 K (B(0)/B(300)) is 1.303. This is the largest for any element. Compared to this, the values for Al, Cu, Fe and Pb are 1.075, 1.036, 1.030 and 1.104 respectively [10]. G(0)/G(300) is also 1.30. This again seems to be unprecedented. In α-Pu phase of Ga alloys, obtained either due to cooling or application of pressure and known as α'-phase, the bulk modulus also softens (40 GPa for 2.3% atm. alloy[11] and 29.1 GPa for 6.55% atm. alloy[12] respectivly).The lattice in these is also expanded.

The pressure derivative of the bulk modulus ( B') of α-Pu has been measured by piston cylinder[13,14] and X-ray diffraction techniques at room temperature[12,3,15]. Table 1 displays these values. All of these are large. For other elements, these are between 4 and 6. Fig.2 compares the mean value of α-Pu from various diffraction measurements with those of the neighbouring actinide elements. Again the value for Pu is extreme. A question arises as to whether it is an artefact due to some experimental conditions or intrinsic. It is well known that a large value may result in an X –ray experiment, if the sample is not under hydrostatic conditions. It may be noted that Dabos-Seigon et al [3] included silicon oil as the pressure transmitting medium and, which is not known to ensure good hydrostatic conditions. Roof did not employ any medium in his diamond cells. However, Faure and Genestier [15] loaded argon. This provides better hydrostatic conditions. Yet, they obtained the highest value for B'. It is also known that, when fitting an equation of state expression to experimental data, strong correlations may be observed *between bulk modulus* and its pressure derivative. Again, Faure and Genestier [15] ruled this out by fits to their data, in which *the ambient pressure volume was fixed and B' varied in a large range*. The resulting bulk modulus values lay in a very narrow range around the value in Table 1.

The values listed by Ledbetter et al of α-Pu for the Gruneisen parameter $\Gamma_0$ range from 3.0 to 9.6[16]. This variation is too large. This value is usually around 2. The quasiharmonic Gruneisen parameter is defined as

$$\Gamma_0 = - < d \ln \omega(V) / d \ln(V) > \qquad (1)$$

where the brackets < > designate an average over all the normal modes of the solid[17]. Gruneisen[18] related it to various thermodynamic quantities by the relation (known as Gruneisen rule)

$$\Gamma_0 = V_0 \alpha B_T / C_V \qquad (2)$$

Where $V_0$ = normal volume, $\alpha$ = coefficient of thermal expansion, $B_T$ = isothermal bulk modulus and $C_V$ = specific heat at constant volume. $\Gamma_0$ is 3.5 for $\alpha$-Pu [16]. $\Gamma_0$ also been related to dB/dP. In Debye- Gruneisen approximation various relations are often used e.g. Donald-MacDonald relation [19].

$$dB/dP = 2\,\Gamma_0 + 2/3 \qquad (3)$$

Table 1 also lists these $\Gamma_0$ values. These differ sharply from that from the Gruneisen expression.

The purpose of this paper is to explain the anomalies in the above physical properties of $\alpha$–Pu, which have existed in literature for almost a decade. Using a self-consistent thermodynamic model, Filantov and Povzer [20] have recently shown that temperature dependence of bulk modulus and thermal expansion are caused by strong lattice anharmonicity in α-Pu rather than by 5f localization-delocalization. We instead show here these are coupled to electronic effects because of temperature changes. As stated above, the atomic volumes and hence thermal expansions of various Pu phases have been related to the 5f correlation length scale (degree of localisation/ delocalisation [5]).The is also applicable for their bulk moduli and shear moduli (see fig.3a). Fig.3b displays this correlation for the $\alpha$ phase alone. Here, the moduli have been taken from Suzuki et al [7] and volumes are from Wallace [21] corresponding to each temperature. The two data for alloys again show a reasonable correspondence (not shown here) though these data are sparse. It may be added here that we find these agreements also for other quantities derived from B and G like Debye temperatures, Young moduli etc. All this is in line with the caveat, well established in mineral physics, attributed to Birch [22], that in a given mineral, the change of volume is the primary agent for the variation of elastic moduli. All other variables like temperature, pressure, composition change, phase changes etc. affect the moduli only as far they affect the volume itself [23]. Below, we use the same law to rationalise the observed high values of the pressure derivative of the bulk modulus[.

We derive this volume variation of the bulk modulus using the well known Vinet equation of state [24].

$$P(V) = 3 B_o X^{-2} (1-X) \exp(3\eta(1-X))) \qquad (4)$$

Where $X = (V/V_o)^{1/3}$ and $\eta = 1.5(B'-1)$. By differentiation with respect to volume, we write

$$B(V) = B_0 X^{-2} \exp(\eta(1-X)) (2-X-X(1-X)\eta) \qquad (5)$$

Comparison of calculated B' values with experimental values for $V_0 = 19.42$ $A^3$/atom at 0K are shown in Table 3. There is good agreement. To rule out the possibility that this agreement is not dependent on the form of the equation of state used, we also employed the expression given by Kunc et al[25] based on equation of state of Aleksandrogv et al[26]. Here

$$B(V) = B_0 (V_o/V) [1 + (B'-1)((V_o/V)-1)] \qquad (6)$$

The values from this are given in column 5 of Table 2. The agreement again shows that the high values for B' at T = 300 K are not fortuitous. The values for ultrasonically measured data at different temperatures are also reasonably reproduced. The dependence of B' on temperature may also be noted. This need to be verified by conducting X- ray experiments at low temperatures.

To determine the most appropriate value of Gruneisen parameter for $\alpha$-Pu, we employ equations (7) and (8), which relate it to the lattice thermal part ($P_{lt}$) of the Mie- Gruneisen equation state [17],

$$P = P_c + P_{lt} \qquad (7)$$

$$P_{lt} = (\Gamma(V)/V) 3 k_B T \text{ for } T > \theta_D \qquad (8)$$

where $\theta_D$ is the Debye temperature. In this, the cold zero degree isotherm, $P_c$ is taken from A.Verma, P. Modak, S.M.Sharma and S.K.Sikka (unpublished), who calculated it by DFT+U method. Here, U is the Hubbard interaction which accounts for the repulsion between f electrons. This has already been successfully used by us [27,28] for Am and Cm, which undergo 5f localisation-delocalisation transition under pressure. The thermal part for T=300 K is then added to this with value of $\Gamma_0$ (under approximation $\Gamma/V = \Gamma_0/V_0$), chosen so that the calculated 300 K isotherm best matches the experimental one. This determined value of $\Gamma_0$ is 3.7. The goodness of fit maybe judged from fitted EOS parameters: V= 19.81A$^3$, B= 68.2 GPa and B'=9.5. These may be compared with the experimental values in Table 1. Our $\Gamma_0$ is the same as recommended by Ledbetter et al[15]. This is also near the value 3.5 obtained by them from Gruneisen rule. For some other actinides $\Gamma_0$ vales are: U (2.29) [18], Np (2.7) Am (0.58) and Cm (0.71) [29,30,31] respectively. The nearest value, 3.04 is that for Au [18]. Filantov and Povzer [18], from their model, get a value of 2.8 at 0 K and 4.0 at 300 K for α-Pu. These bracket our determined value. A temperature dependence of the pressure derivative has been noted in Table 2. This will also be reflected in Γ values. It is interesting to point out that at 96 K (see Table 2); its predicted value is 3. Thus, we agree with Ledbetter et al [15] that their high Gruneisen parameters may be temperature induced artefact.

In summary, for α Pu, we have related (i) the elastic moduli dependence on temperature with iterant to localization transition of 5f electrons, (ii) large value of the pressure derivative of bulk modulus to its interplay with volume and bulk modulus and (iii) determined the appropriate value of Gruneisen parameter through equation of state calculations.

I thank Dr. A.Verma, Dr. P.Modak and Dr. S.M. Sharma for useful discussions.

Table 1 Experimental data for bulk modulus and its pressure derivative of α-Pu at 300 K. The derived Gruneisen parameters from Dugdale MacDonalad expression (eq.3) are also shown.

| Method | V/atom (A$^3$) for T=300 K | Bulk modulus –B (GPa) | Pressure derivative-B' | Gruneisen parameter- $\Gamma_0$=(B'-2/3)/2 | Reference |
|---|---|---|---|---|---|
| Piston cylinder | not available | 51(2) | 12.1 | 5.7 | [13] |
| Piston cylinder | 20.07 | 47.6 | 14 | 6.7 | [14] |
| X-ray | 20.06(6) | 42 | 10.5 | 4.9 | [12] |
|  |  | 40 | 14 | 6.2 | Vinet [] fit to the same data |
| X-ray | 19.88 | 42(2) | 15 | 7.2 | [3] |
| X-ray | 20.08 | 37(2) | 19 | 9.2 | [15] |
| X-ray | 20.32 | 40 | 9 | 4.2 | [11] for Pu-2.3% atm. Ga alloy |
| X-ray | 21.2 | 26 | 10.1 | 4.9 | [12] for Pu-6.5% atm. Ga alloy |

Table 2. Comparsion of calculated pressure derivatives for α-Pu from quations 7 and 8 with experimental data as a function of temperature and $V_0$ = 19.4 $A^3$ .

| V/atom($A^3$) | T (K) | B (GPa) | B' from eg.7 | B' from eq.8 | B'-expertmental |
|---|---|---|---|---|---|
| 19.87 | 300 | 41.3 | 20.5 | 17.6 | 15(2) [3] |
| 20.07 | 300 | 47.6 | 11.3 | 10.1 | 14 [14] |
| 20.06 | 300 | 39.8 | 16 | 14.2 | 14(1) [12] |
| 20.08 | 300 | 37 | 17.7 | 15.0 | 19.5(5)[15] |
| 20.32 | 300 | 40 | 8.9 | 8.2 | 9(2) [11] |
| 21.20 | 300 | 26.1 | 9.6 | 8.1 | 10.1(5) [7] |
| 20.28 | 380 | 47 | 8.9 | 8.2 | -   [7] |
| 319.98 | 300 | 54 | 9 | 8.4 | -   [7] |
| 19.7 | 220 | 61 | 7.8 | 7.7 | -   [7] |
| 19.54 | 96 | 68 | 6.9 | 6.5 | -   [7] |

Fig.1 Temperature dependence of elastic moduli of α-Pu. The data is taken from Suzuki et al [7].

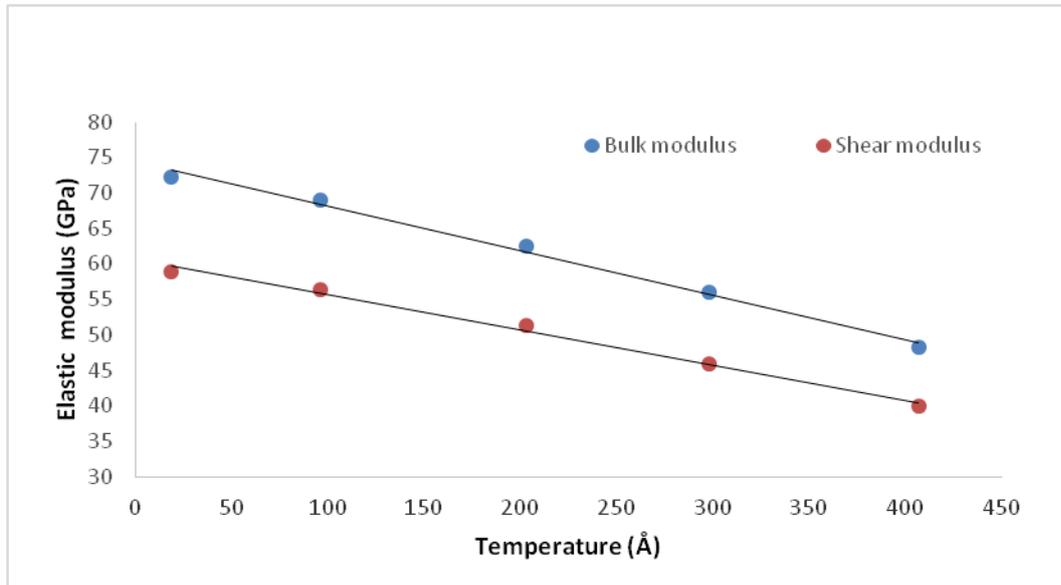

Fig.2 Pressure derivatives of the bulk modulus of actinides showing the anomalously large value for α-Pu.

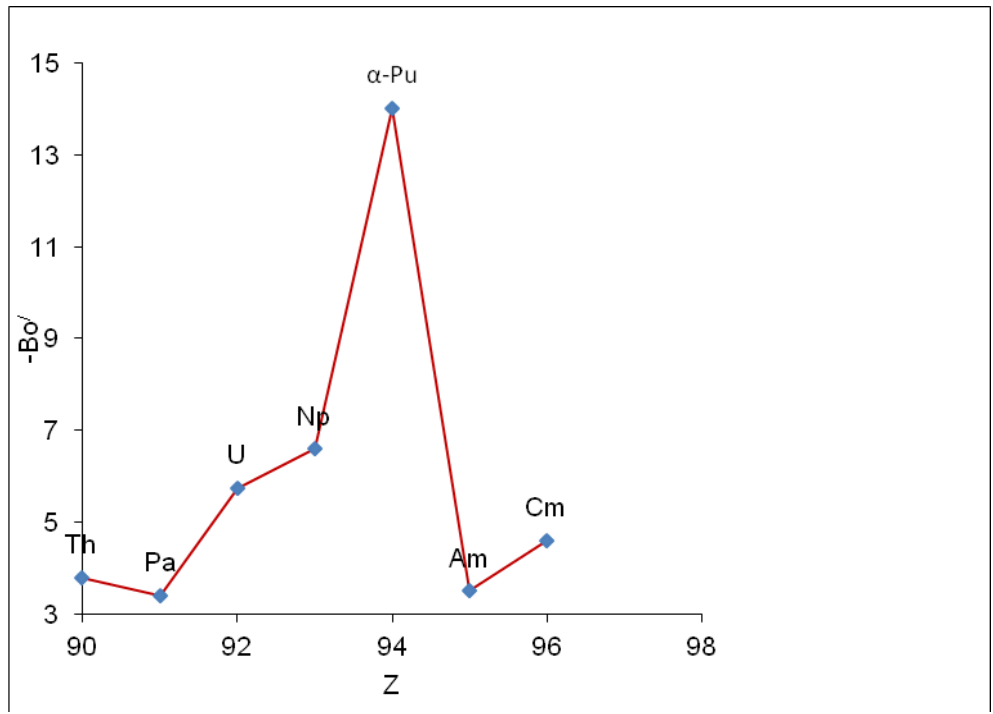

Fig.3 Bulk and shear moduli of (a) different temperature induced pu phases and of (b) α-pu alone as a function of temperature versus 5f correlation length defined by Savane et al [5]. C=0 represents fully delocalized 5f and C=1 fully localized 5f electrons.

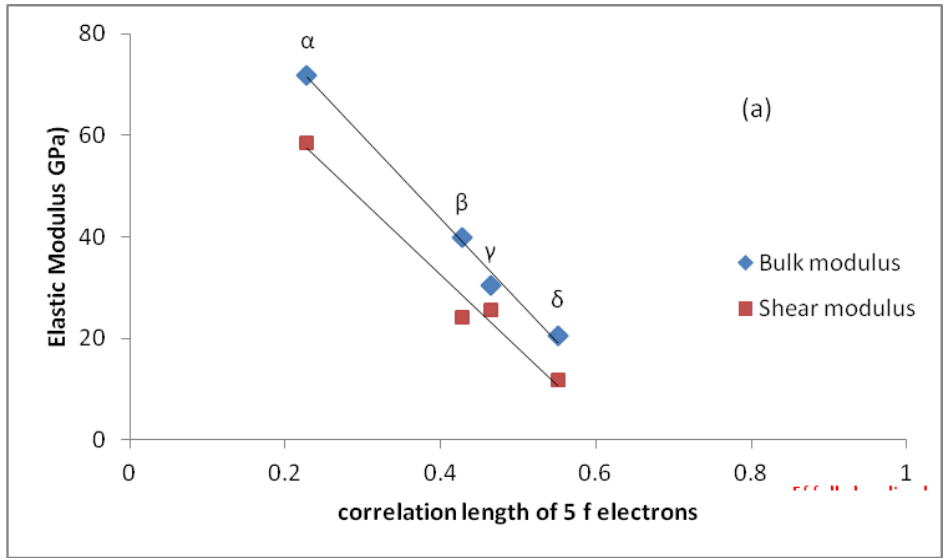

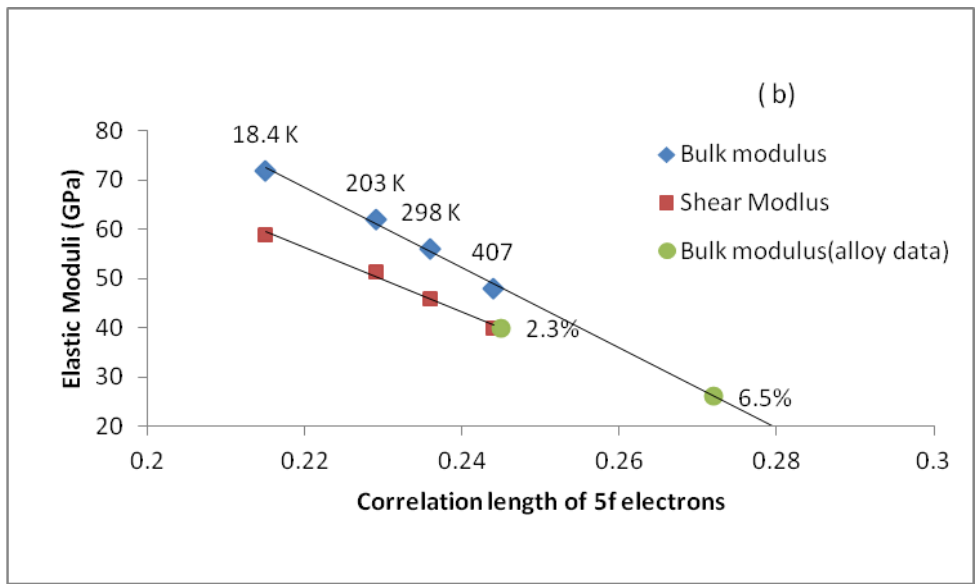

Fig.4 Comparison of the experimental and calculated 300 K isotherm For α-Pu, the calculated one is for $\Gamma_0 =3.7$.

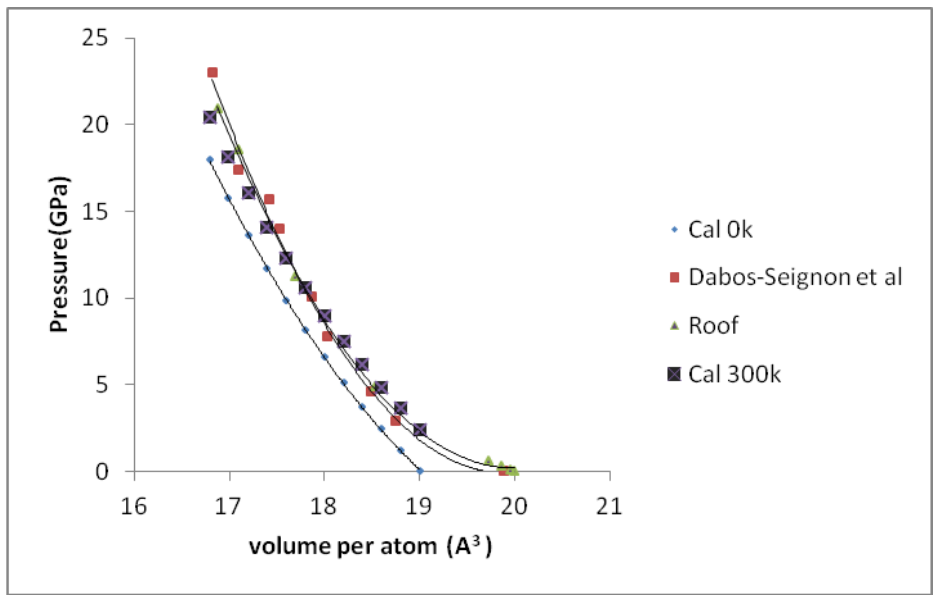